\documentclass{epl}

\usepackage{graphicx}
\usepackage{hyperref}
\hypersetup{%
  pdftitle={Decoherence of spatially separated qubits},%
  pdfauthor={Roland Doll, Martijn Wubs, Peter H\"anggi, Sigmund Kohler},%
  pdfkeywords={quantum information, qubits, spatial separation, entanglement,%
               decoherence, decoherence-free subspace},%
  pdfpagemode=None}

\usepackage{amsmath}
\usepackage{amssymb}
\usepackage{units}
\usepackage{wasysym}

\newcommand{\im}{\mathrm{i}}
\newcommand{\diff}{\mathrm{d}}
\newcommand{\e}{\mathrm{e}}
\DeclareMathOperator{\trace}{tr}
\DeclareMathOperator{\Real}{Re}
\DeclareMathOperator{\Imag}{Im}

\title{Limitation of entanglement due to spatial qubit separation}

\author{Roland Doll\thanks{E-mail: \email{roland.doll@physik.uni-augsburg.de}}
\and Martijn Wubs \and Peter H\"anggi \and Sigmund Kohler}
\shortauthor{R. Doll \etal}

\institute{%
Institut f{\"u}r Physik, Universit{\"a}t Augsburg, Universit\"atsstra{\ss}e 1,\\
D-86135 Augsburg, Germany}


\begin{document}

\maketitle
\begin{abstract}
We consider spatially separated qubits coupled to a thermal bosonic field that
causes pure dephasing. Our focus is on the entanglement of two Bell states which
for vanishing separation are known as robust and fragile entangled states. The
reduced two-qubit dynamics is solved exactly and explicitly. Our results allow
us to gain information about the robustness of two-qubit decoherence-free
subspaces with respect to physical parameters such as temperature, qubit-bath
coupling strength and spatial separation of the qubits. Moreover, we clarify the
relation between single-qubit coherence and two-qubit entanglement and identify
parameter regimes in which the terms robust and fragile are no longer
appropriate.
\end{abstract}

In recent years, we witnessed a fast development of the experimental realization
of quantum bits and the maintenance of their quantum coherence.  Despite this
progress, decoherence stemming from a coupling of the qubits to a macroscopic
environment remains a major obstacle for the implementation of a quantum
computer.
Several strategies are pursued to beat decoherence. One is dynamical decoupling:
single qubits~\cite{Viola1998a} or two-qubit gates~\cite{FonsecaRomero2005a} are
effectively isolated from their environment by driving them with ac fields.
Another strategy is quantum error correction, which requires a redundant
encoding of a logical qubit by several physical qubits. Standard error
correction protocols presuppose that all physical qubits couple to uncorrelated
baths~\cite{Unruh1995a,DeChiara2004a,Terhal2005a}, which can be realized by
putting qubits far apart~\cite{Duan1998a}. 
A third, more direct strategy is the use of a decoherence-free subspace (DFS).
There, one logical qubit is encoded by several physical qubits, in such a way
that the logical qubit states do not couple to the environment~\cite{Palma1996a,
Zanardi1997a, Lidar1998a}. Ideal DFSs occur when physical qubits couple via a
collective coordinate to a common bath and are fairly robust against
perturbations~\cite{Bacon1999a, Altepeter2004a}.

In view of the above, it is an important question whether different qubits are
exposed to spatially correlated or uncorrelated noise.  For example for charge
qubits in quantum dots~\cite{Hayashi2003a}, a relevant source of decoherence is
the coupling to substrate phonons~\cite{Storcz2005a,Thorwart2005a}. Widely
separated charge qubits at high temperatures will experience uncorrelated noise
due to the phonon bath.  The identification of a crossover regime at lower
temperatures and smaller separations in which spatial correlations become
important, requires a model in which the qubit separation and the phonon bath
are taken explicitly into account.

In this Letter, we study the decoherence of spatially separated qubits which are
coupled to a bosonic field that causes phase noise. On short time scales, phase
noise is the main source of decoherence in solid state
environments~\cite{Makhlin2001a}; on longer time scales, bit-flip noise becomes
relevant as well~\cite{Mozyrsky2002a}, but we focus on the former herein. Phase
noise is characterised by a qubit-bath coupling that commutes with the qubit
Hamiltonian and, thus, allows an exact solution of the dissipative quantum
dynamics. Still, the evaluation of the resulting exact expressions can be rather
complex and often relies on approximations~\cite{Unruh1995a, Palma1996a,
Duan1998a, Mozyrsky1998a, Yu2002a, Yu2003a, Reina2002a, Tolkunov2005a,
Braun2006a}.
In contrast to those works, we here evaluate the exact solution in explicit
form. We focus on the entanglement of two qubits prepared in particular Bell
states and derive explicit expressions showing how their entanglement changes
upon increasing their separation. For a wide range of parameters, we find that
the dynamics is highly non-Markovian and that the entanglement can converge to
relatively large values, even at high temperatures. We will discuss the
robustness of the two-qubit DFS and clarify the relation between two-qubit
entanglement and single-qubit coherence.

\section{Robust and fragile entangled states}

Yu and Eberly~\cite{Yu2002a} studied the entanglement dynamics for two qubits
coupled to the same heat bath at \textit{identical} positions, in particular for
a preparation of the (maximally entangled) Bell states
\begin{equation}
\label{eqBellStates}
|{\psi_\mathrm{robust}}\rangle =|{\psi_{-}}\rangle
= \frac{|{01}\rangle + |{10}\rangle}{\sqrt{2}}\,,
\qquad
|{\psi_\mathrm{fragile}}\rangle = |{\psi_{+}}\rangle
= \frac{{|00}\rangle + |{11}\rangle}{\sqrt{2}}\,,
\end{equation}
where $|{\mathbf{n}}\rangle = |{n_{1},n_{2},\ldots, n_{N}}\rangle$ denotes the
$N$ qubit state with $\sigma_{\nu z}|{\mathbf{n}}\rangle =
(-1)^{n_\nu}|{\mathbf{n}}\rangle$, $n_\nu=0,1$ and $\sigma_{\nu z}$ a Pauli
matrix for qubit~$\nu$.
They found that the ``robust state'' lives in a DFS and consequently its
entanglement is preserved, whereas the entanglement of the qubits initially in
the ``fragile state'' decays to zero.  In our study, we will also consider these
initial states. The notation $|{\psi_{\pm}}\rangle$ has been introduced for
writing equations more efficiently.

Two-qubit entanglement can be measured by the concurrence $C[\rho] =
\max\{0,\sqrt{\lambda_1} -\sqrt{\lambda_2} -\sqrt{\lambda_3}
-\sqrt{\lambda_4}\}$ where $\lambda_i$ denotes the eigenvalues of the matrix
$\rho \sigma_{1y}\sigma_{2y} \rho^\ast \sigma_{1y}\sigma_{2y}$ in decreasing
order;  $\rho^\ast$ is the complex conjugate of $\rho$ \cite{Wootters1998a}. For
maximally entangled states, one finds $C=1$, while $C$ vanishes for incoherent
mixtures of product states. It was noticed~\cite{Yu2002a} that for phase noise
and two qubits at vanishing separation initially in either of the states
\eqref{eqBellStates}, the concurrence is given by the absolute values of
particular density matrix elements  $\rho_{\mathbf{m},\mathbf{n}} = \langle
\mathbf{m}|\rho|\mathbf{n}\rangle$, namely  $C_- = 2|\rho_{01,10}|$ and $C_+ =
2|\rho_{00,11}|$, respectively. These relations hold for spatially separated
qubits as well.

\section{Qubits coupled to a bosonic field}

For modelling $N$ identical qubits coupled to a homogeneous bosonic environment,
we employ a spin-boson Hamiltonian
\begin{equation}
\label{eqH}
H = \frac{\Delta}{2} \sum_{\nu = 1}^N \sigma_{\nu z}
   +\sum_\mathbf{k} \hbar \omega_k b_\mathbf{k}^\dagger b_\mathbf{k}
   + H_\text{q-b}\,.
\end{equation}
The first term denotes $N$ qubits with energy splittings $\Delta$ and the second
term represents a bosonic field with isotropic dispersion relation
$\omega_\mathbf{k}=\omega_{k}= c|\mathbf{k}|$ and sound velocity $c$. The
qubit-bath interaction
\begin{equation}\label{Hqb}
H_\text{q-b} = \hbar \sum_{\nu=1}^{N}\sigma_{\nu z}\xi_{\nu},
\quad \text{with}\quad
\xi_\nu = \sum_\mathbf{k} g_k
            (b_\mathbf{k} \e^{\im \mathbf{k}\cdot\mathbf{x}_\nu}
            +b_\mathbf{k}^\dagger \e^{-\im \mathbf{k}\cdot\mathbf{x}_\nu}
            ) ,
\end{equation}
introduces phase noise due the coupling of qubit $\nu$ via $\sigma_{\nu z}$ to
the field at position $\mathbf{x}_\nu$. We assumed identical and isotropic
coupling strengths for each qubit, i.e.\ $g_{\mathbf{k}\nu} = g_{k}$.

Furthermore, we assume that at initial time $t=0$, the density matrix of qubit
plus bath, $R(0)$, is of the Feynman-Vernon type, i.e.\ the bath is in thermal
equilibrium and uncorrelated with the qubits, $R(0) = \rho(0) \otimes
\rho_\mathrm{b}^\mathrm{eq}$, where $\rho_\mathrm{b}^\mathrm{eq} \propto
\exp\{-\sum_\mathbf{k} \hbar\omega_\mathbf{k} b_\mathbf{k}^\dagger b_\mathbf{k}
/ k_\mathrm{B}T\}$ is the canonical ensemble of the bosons.  Since we are
exclusively interested in the behaviour of the qubits, we trace out the bath and
obtain the qubits' reduced density operator $\rho(t) = \trace_\mathrm{b} R(t)$.
From the Liouville-von Neumann equation in the usual interaction picture,
$\im\hbar\,\diff\widetilde R/\diff t = [\widetilde H_\text{q-b}(t), \widetilde
R]$, we obtain after some algebra for the density matrix elements the closed
expression~\cite{Palma1996a,Duan1998a,Reina2002a}
\begin{equation} \label{eqExactResult}
  \tilde\rho_{\mathbf{m},\mathbf{n}}(t)
=
  \rho_{\mathbf{m},\mathbf{n}}(0)\,\e^{-\Lambda_{\mathbf{m},\mathbf{n}}(t)
  + \im \phi_{\mathbf{m},\mathbf{n}}(t)}\,.
\end{equation}
The phases $\phi_{\mathbf{m},\mathbf{n}}(t)$ correspond to Lamb shifts which are
brought about by time ordering.  Since all quantities considered in the
following are given by absolute values of single density matrix elements, these
phases will not be relevant. The amplitude damping depends on the qubit
separations
$\mathbf{x}_{\nu\nu^\prime} = \mathbf{x}_\nu -
\mathbf{x}_{\nu^\prime}$ and reads
\begin{equation} \label{eqDamping}
\Lambda_{\mathbf{m},\mathbf{n}}(t)
= \sum_\mathbf{k} g_k^2 \frac{1-\cos(\omega_k t)}{\omega_k^2}
   \coth\left( \frac{\hbar \omega_k}{2k_\text{B}T}\right)
   \Bigg| \sum_{\nu,\nu^\prime=1}^N
          [(-1)^{m_\nu} -(-1)^{n_{\nu}}]
          \e^{-\im\mathbf{k}\cdot \mathbf{x}_{\nu\nu^\prime}}
   \Bigg|^2 .
\end{equation}

For the evaluation of these damping factors, it is convenient to introduce the
spectral density $J^{(d)}(\omega) \equiv \sum_\mathbf{k} g_{k}^{2}
\delta(\omega-ck)$.  For an homogeneous isotropic $d$-dimensional
environment, it reads $J^{(d)}(\omega) = \alpha \omega
(\omega/\omega_\mathrm{c})^{d-1} \exp\{-\omega/\omega_\mathrm{c}\}$~\cite{Weiss1999a},
featuring the damping strength $\alpha$ and the cutoff frequency
$\omega_\mathrm{c}$ which for phonons is the Debye frequency.
Then, the summation over the wave vectors $\mathbf{k}$ can be replaced by a
frequency integration plus an integration over the solid angle. Evaluating the
latter, we obtain for the concurrence the exact expression
\begin{align}
\label{eqCExactImplicit}
C^{(d)}_\pm(t) &= \exp\biggl\{-8 \int_0^\infty\diff\omega\,
J^{(d)}(\omega)\biggl[1 \pm G^{(d)} \left(\frac{\omega x_{12}}{c}\right) \biggr]
\frac{1-\cos(\omega t)}{\omega^2}
\coth\left(\frac{\hbar\omega}{2k_\mathrm{B} T}\right)\biggr\} ,
\end{align}
where $G^{(d)}(x)$ denotes a dimension-dependent geometrical factor
which reads $G^{(1)}(x) = \cos(x)$, $G^{(2)}(x) = J_0(x)$, and
$G^{(3)}(x) = \sin(x) / x$, with $J_0$ the zeroth-order Bessel
function of the first kind.
Note that $G^{(d)}(0) = 1$ for all dimensions, which causes the robust and the
fragile behaviour of the Bell states~\eqref{eqBellStates} for vanishing
separation. For large argument, $G^{(d)}$ decays for $d=2,3$, but not for $d=1$.

Below, we will compare the concurrence of a qubit pair with the single-qubit
coherence $|\rho_{0,1}(t)/\rho_{0,1}(0)|=\exp\{-\Lambda^{(d)}_{0,1}(t)\}$, which
we define for $N\!=\!1$, i.e.\ when only one qubit is present. It is formally
given by the rhs of eq.~\eqref{eqCExactImplicit} but with the replacement $1\pm
G^{(d)} \to \frac{1}{2}$. Thus it is the geometrical factor $G^{(d)}$ which
determines the difference between single-qubit decoherence and entanglement
decay of a qubit pair. In particular, the dimension dependence of $G^{(d)}$ will
turn out to be crucial but has been ignored in prior studies~\cite{Reina2002a}.
For qubits coupled to a three-dimensional bath, we find in the remote limit
$x_{12}\to\infty$ for both concurrences the relation $C^{(3)}_{\pm}(t)=
\exp\{-2\Lambda^{(3)}_{0,1}(t)\}$, which was also obtained for a model
consisting of independent baths \cite{Tolkunov2005a}.  More generally, an
intriguing corollary to eq.~\eqref{eqCExactImplicit} is the exact relation
$C^{(d)}_{+}(t)\, C^{(d)}_{-}(t) = \exp\{-4\Lambda^{(d)}_{0,1}(t)\}$, which for
arbitrary separation links the concurrences to the single-qubit coherence.  It
implies that if one of the concurrences vanishes, the single-qubit coherence
must vanish as well.  A finite single-qubit coherence, in turn, requires
non-vanishing concurrences.

In order to evaluate the concurrences \eqref{eqCExactImplicit}, we introduce the
scaled time $\tau = \omega_\mathrm{c}t$ and transit time $\tau_{12} =
\omega_\mathrm{c}x_{12}/c$, and the scaled temperature $\theta = k_\mathrm{B} T/
\hbar\omega_\mathrm{c}$. After inserting a Taylor expansion for $\cos(\omega
\tau)$, we accomplish the frequency integrals in the resulting
series~\cite{GradshteynRyzhik_ssq}. Then we obtain for the single-qubit
coherence and likewise for the concurrence an infinite product which can be
combined into gamma functions and their derivatives. We restrict ourselves to
the cases of one and three dimensions and find for the former case the
single-qubit coherence
\begin{equation}\label{eqSingleQubitDecoherence1d}
\e^{-\Lambda^{(1)}_{0,1}(\tau)}
= \left| \frac{\Gamma(\theta[1+\im \tau])
 \Gamma(\theta[1-\im \tau])}{\Gamma^2(\theta)}\right|^{4\alpha}
 \left(1+\tau^2\right)^{2\alpha}\,,
\end{equation}
and the concurrence
\begin{equation} \label{eqC1(t)}
C^{(1)}_\pm(\tau)
= \e^{-2\Lambda^{(1)}_{0,1}(\tau)}
  \left| \frac{\Gamma(\theta[1-\im(\tau_{12} - \tau)])
  \Gamma(\theta[1-\im(\tau_{12} + \tau)])}{\Gamma^2(\theta[1-\im\tau_{12}])}
  \right|^{\pm8\alpha} \left|1+\frac{\tau^2}
  {(1-\im\tau_{12})^2}\right|^{\pm 4 \alpha}\,,
\end{equation}
where $\Gamma$ denotes the Euler gamma function.
The corresponding expressions for a three-dimensional environment read
\begin{align}
\label{eqSingleQubitDecoherence3d}
\e^{-\Lambda^{(3)}_{0,1}(\tau)}
&= \exp\left\{ -
   4 \alpha \bigg(
   2\theta^2 \Real \bigl[\,\Psi_1(\theta) - \Psi_1(\theta[1-\im \tau])\,\bigr] -
   \frac{\tau^2(\tau^2 + 3)}{(1+\tau^2)^2}\bigg)\right\}\,, \\ \label{eqC3(t)}
\begin{split}
C_\pm^{(3)}(\tau)
&= \e^{-2\Lambda^{(3)}_{0,1}(\tau)}
   \exp\bigg\{ \pm 8 \alpha \bigg( \frac{\theta}{\tau_{12}} \Imag
   \bigl[\, 2 \Psi_0(\theta[1-\im \tau_{12}]) - \Psi_0(\theta[1-\im(\tau-\tau_{12})]) \\
& \qquad - \Psi_0(\theta[1-\im(\tau+\tau_{12})]) \, \bigl] +
  \frac{\tau^2(\tau^2-\tau_{12}^2 + 3)}{(1+\tau_{12}^2)(\tau^4
  -2\tau^2\left[ \tau_{12}^2-1\right] +\left[1+\tau_{12}^2\right]^2)}
  \bigg)\bigg\}\,,
\end{split}
\end{align}
where $\Psi_0$ and $\Psi_1$ are Di-Gamma and Tri-Gamma functions, respectively. 
The importance of eqs.~\eqref{eqC1(t)} and \eqref{eqC3(t)} lies in the fact that
they explicitly yield the concurrences at all times for arbitrary spatial
separations $\tau_{12}$, from a perfect DFS ($\tau_{12}=0$) to uncorrelated
noise ($\tau_{12} \rightarrow \infty$). In both expressions, the respective
single-qubit coherences \eqref{eqSingleQubitDecoherence1d} and
\eqref{eqSingleQubitDecoherence3d} appear. 

\section{Time-evolution of the robust Bell state}

\begin{figure}
\twoimages{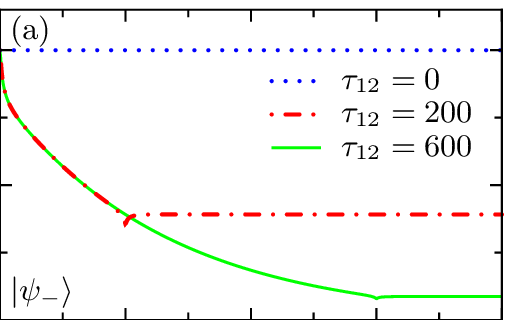}{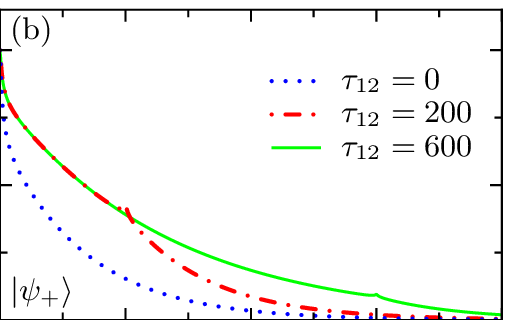}
\twoimages{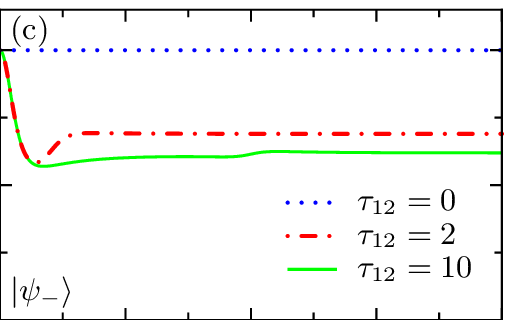}{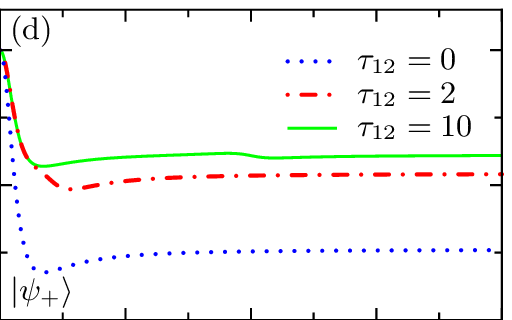}
\caption{\label{figConcurrenceExact}
Time evolution of the concurrence for the robust (left) and the
fragile (right) Bell state for various spatial separations
$c\tau_{12}/\omega_\mathrm{c}$. The qubits
couple with a strength $\alpha = 0.01$ to a one-dimensional (a,b) and a
three-dimensional bath (c,d), respectively, at temperature $\theta
=k_\mathrm{B}T/\hbar\omega_\mathrm{c}=0.015$. For qubit energies $\Delta =
0.01\hbar\omega_\mathrm{c}$, the time range in the upper (lower) plots
corresponds to 1.3 (0.03) coherent oscillations.}
\end{figure}
Let us first focus on the entanglement of a qubit pair that starts out in the
robust state $|{\psi_-}\rangle$ and couples to a one-dimensional heat
bath. Figure \ref{figConcurrenceExact}a depicts the time-evolution of the
concurrence for a temperature well below the Debye temperature. For vanishing
separation, $\tau_{12}=0$, the concurrence $C^{(1)}_{-}(\tau)$ remains at its
initial value 1.  This reflects the fact that then $|\psi_-\rangle$ lives in a
DFS and, consequently, is robust. For $\tau_{12}>0$, we find that the
concurrence initially decays until the transit time $\tau_{12}$ is reached. At
time $\tau=\tau_{12}$, the decay comes to a standstill and the concurrence
remains at a finite value $C_-^{(1)}(\tau\to\infty) =
(1+\tau^2_{12})^{4\alpha}\left|\Gamma(\theta[1-\im\tau_{12}])/\Gamma(\theta)\right|^{16\alpha}$
which becomes $(1+\tau^2_{12})^{-4\alpha}$ for $\theta\to 0$. The time evolution
allows the interpretation that before the transit time is reached,
uncorrelated noise affects the qubits and entails an entanglement
decay. After the transit time, the noise at the two positions is
sufficiently correlated to establish a
decoherence-free subspace. In the remote limit $\tau_{12}\to\infty$, the
concurrence of the robust state finally vanishes and, thus, the residual
entanglement for finite $\tau_{12}$ can be attributed to spatial bath
correlations.  We emphasise that for a one-dimensional bath, this
scenario holds true for all temperatures.

An intuitive physical picture for the observed entanglement dynamics
is that at long times, decoherence is governed by the low-frequency
modes of the bath. Owing to their large wavelengths, these modes act
effectively as a collective bath coordinate which leaves
the entanglement robust. In higher dimensions, the role of the
low-frequency modes is suppressed and, thus, the long-time behaviour
may be significantly different~\cite{Palma1996a}.

\begin{figure}
\twofigures{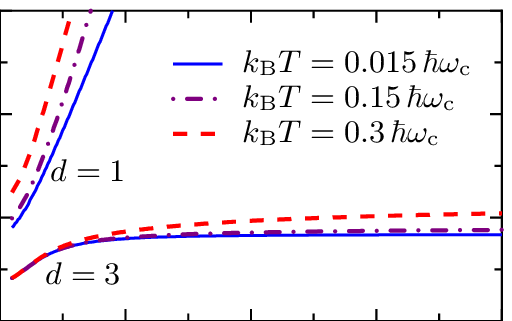}{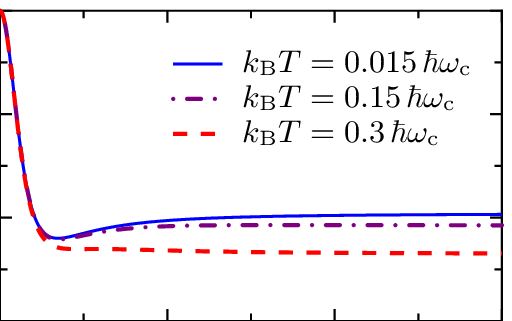}
\caption{\label{figDecayTime}
Duration $\tau^\ast$ of the concurrence decay of the robust Bell state
$|\psi_-\rangle$ as a function of the transit time $\tau_{12}$
for one- and three-dimensional baths. The dissipation strength is $\alpha = 0.01$.}
\caption{\label{figSingleQubitDecoherence} Time evolution of the
single-qubit coherence for various temperatures. The qubit is coupled with
the strength $\alpha = 0.01$ to a three-dimensional bath.}
\end{figure}

Figure \ref{figConcurrenceExact}c 
reveals that for a three-dimensional bath, in general the concurrence~\eqref{eqC3(t)} decays and saturates
at a finite value which stays larger for closer qubits. In the
experimentally relevant limit of low temperatures and
$\tau_{12}\theta^2 \apprle 1$, the final concurrence emerges as
\begin{equation}
\label{eqC3-(infty)}
C_-^{(3)}(\tau\to\infty) = \exp \biggl\{ - 8 \alpha \biggl( 1 -
\frac{1}{1+\tau_{12}^2} + \frac{\pi^4\tau_{12}^2\theta^4}{45} \biggr)\biggr\}\,.
\end{equation}
However, the saturation generally occurs already at a time $\tau \ll
\tau_{12}$, i.e.~long before a field distortion can have propagated
from one qubit to the other. To make this statement
more quantitative, we numerically estimate the duration $\tau^\ast$ of the
concurrence decay by the time at which 90\% of the decay has happened.
Figure~\ref{figDecayTime} shows that $\tau^\ast \approx 1$.  In particular,
$\tau^\ast$ is independent of the spatial separation $\tau_{12}$, unless the
qubits are very close. Hence the saturation of $C^{(3)}_{-}(\tau)$ cannot be
explained as a delayed build-up of a DFS. Instead, a single-qubit mechanism must
be at work, since at times $\tau < \tau_{12}$, the qubits experience effectively
\textit{uncorrelated} noise.  This conjecture is supported by the resemblance of
$C_-^{(3)}(\tau)$ to the single-qubit coherence $\exp \{
-\Lambda^{(3)}_{0,1}(\tau) \}$ shown in fig.~\ref{figSingleQubitDecoherence}.
A second difference to the one-dimensional case concerns the remote
limit of the qubits: For $\tau_{12}\to\infty$, the stationary value is
still finite. In this limit, $G^{(3)}$ in eq.~(\ref{eqCExactImplicit})
is negligible and the concurrence is given by the square of the finite
single-qubit coherence.

The concurrence~\eqref{eqC3-(infty)} depends only quadratically on $\tau_{12}$
and, thus, is rather robust against variations of the separation, provided the
separation is small. Such robustness was predicted~\cite{Bacon1999a} and
confirmed experimentally~\cite{Altepeter2004a} for symmetry-breaking
perturbations. Interestingly, eq.~\eqref{eqC3-(infty)} shows that the
concurrence is robust against temperature variations as well, about which the
theory in ref.~\cite{Bacon1999a} makes no predictions.

\section{Time-evolution of the fragile Bell state}

If the qubits are initially in the fragile state $|\psi_+\rangle$, a
one-dimensional bath causes an entanglement decay that becomes faster
once the transit time is reached; see fig.~\ref{figConcurrenceExact}b.
As for the robust state $|\psi_-\rangle$, cooperative effects
only set in after a time $\tau_{12}$. Whether the qubits are spatially
separated or not, for a one-dimensional environment their concurrence
$C^{(1)}_{+}(\tau)$ ultimately decays to zero. The
three-dimensional case again bears more surprises, as seen in
fig.~\ref{figConcurrenceExact}d: the concurrence~\eqref{eqC3(t)}
decays, but the final value $C_{+}^{(3)}(\tau\to\infty)$ is 
nonzero, in contrast to earlier statements~\cite{Yu2002a,Yu2003a}.
For low temperatures such that $\theta \apprle 1,\tau_{12}^{-1/2}$, the
long-time limit is given by
\begin{equation} \label{eqC3+(infty)}
C_+^{(3)}(\tau\to\infty) 
= \exp \biggl\{ - 8 \alpha \biggl( 1
  + \frac{1}{1+\tau_{12}^2} + \frac{2\pi^2\theta^2}{3} \biggr)\biggr\}\, .
\end{equation}
This asymptotic value can be increased by reducing the temperature and by
increasing the qubit separation.  For a separation $\tau_{12}\apprge
1$, $C_+^{(3)}(\infty) = C_-^{(3)}(\infty)$, i.e.\ the concurrence of
both the ``robust'' and the ``fragile'' state become identical; cf.\
the solid lines in figs.~\ref{figConcurrenceExact}c,d.

\section{Conclusions}

The explicit evaluation of the exact reduced dynamics of two qubits with a
nondemolitian coupling to a bosonic heat bath allowed us to investigate the
consequences of a spatial qubit separation.  We focused on two Bell states whose
entanglement for vanishing separation is either robust or fragile.
The most significant consequence of a finite spatial separation is that the entanglement
of the robust Bell state no longer remains robust: it decays initially, yet
after a time $t^\ast = \tau^\ast/\omega_\mathrm{c}$, it saturates. Indeed, it is
interesting to find  stable finite bipartite entanglement even at high
temperatures in our macroscopic system-bath model, which resembles recent
results for a quite different model~\cite{Ferreira2006a}. The duration $t^\ast$
of the entanglement decay depends sensitively on the dimension of the
environment: In one dimension, it equals the transit time of the field from one
qubit to the other. In three dimensions, by contrast, the saturation is governed
by a single-qubit effect and $t^\ast$  is approximately given by the inverse of
the cutoff frequency $\omega_\mathrm{c}$.  Rather surprisingly, these durations
exhibit only a weak temperature dependence.

In a typical solid-state substrate, the Debye temperature is of the order
$\unit[500]{K}$ which corresponds to $\hbar\omega_\mathrm{c} =\unit[40]{meV}$.
For a qubit separation of $\unit[300]{nm}$ and a sound velocity $c
=\unit[3000]{m/s}$, we find for a one-dimensional environment $t^\ast =
\tau^\ast/\omega_\mathrm{c} \approx\unit[10^{-10}]{s}$ while in three
dimensions, this time scale is much shorter: $t^\ast \approx\unit[10^{-13}]{s}$.
For a typical tunnel splitting $\Delta = \unit[10]{\mu eV}$, the coherent
oscillation period is $2\pi\hbar/\Delta \approx \unit[10^{-10}]{s}$, so that in
the three-dimensional environment the concurrence of the robust Bell state
undergoes a decay only during a very short initial stage.  At later times, a
decoherence-free subspace is established and, thus, the concurrence stays
robust.

For the fragile Bell state, a three-dimensional environment in combination with
a finite qubit separation prevents the entanglement from decaying entirely.  If
the qubits are sufficiently well separated, i.e.\ for $x_{12} \apprge
c/\omega_\mathrm{c}$,  the entanglement of the ``fragile'' and the ``robust''
Bell state even assumes practically the same final value.
In the above example, this is already the case if the qubit-qubit distance is
larger than $\unit[1]{\mu m}$, which usually holds for solid-state qubits.  For
typical parameters, the concurrence initially drops to and then remains at values
of the order 0.9 already long before a first coherent oscillation is performed.
Thus, uncorrelated phase noise creates decoherence-poor subspaces, which might
be used for quantum information processing when complemented with quantum error
correction protocols.

\acknowledgements
This work was supported by DFG through SFB 484 and SFB 631.




\begin{thebibliography}{10}

\bibitem{Viola1998a}
\Name{Viola~L. \and Lloyd~S.}
\REVIEW {Phys. Rev. A}{58}{1998}{2733}.

\bibitem{FonsecaRomero2005a}
\Name{Fonseca-Romero~K.~M., Kohler~S., \and H\"anggi~P.}
\REVIEW {Phys. Rev. Lett.}{95}{2005}{140502}.

\bibitem{Unruh1995a}
\Name{Unruh~W.~G.}
\REVIEW {Phys. Rev. A}{51}{1995}{992}.

\bibitem{DeChiara2004a}
\Name{Chiara~G.~D., Fazio~R., Macchiavello~C., \and Palma~G.~M.}
\REVIEW {Europhys. Lett.}{67}{2004}{714}.

\bibitem{Terhal2005a}
\Name{Terhal~B.~M. \and Burkard~G.}
\REVIEW {Phys. Rev. A}{71}{2005}{012336}.

\bibitem{Duan1998a}
\Name{Duan~L.-M. \and Guo~G.-C.}
\REVIEW {Quantum Semiclass. Opt.}{10}{1998}{611}.

\bibitem{Palma1996a}
\Name{Palma~G.~M., Suominen~K.-A., \and Ekert~A.~K.}
\REVIEW {Proc. R. Soc. London, Ser. A}{452}{1996}{567}.

\bibitem{Zanardi1997a}
\Name{Zanardi~P. \and Rasetti~M.}
\REVIEW {Phys. Rev. Lett.}{79}{1997}{3306}.

\bibitem{Lidar1998a}
\Name{Lidar~D.~A., Chuang~I.~L., \and Whaley~K.~B.}
\REVIEW {Phys. Rev. Lett.}{81}{1998}{2594}.

\bibitem{Bacon1999a}
\Name{Bacon~D., Lidar~D.~A., \and Whaley~K.~B.}
\REVIEW {Phys. Rev. A}{60}{1999}{1944}.

\bibitem{Altepeter2004a}
\Name{Altepeter~J.~B., Hadley~P.~G., Wendelken~S.~M., Berglund~A.~J., \and
  Kwiat~P.~G.}
\REVIEW {Phys. Rev. Lett.}{92}{2004}{147901}.

\bibitem{Hayashi2003a}
\Name{Hayashi~T., Fujisawa~T., Cheong~H.~D., Jeong~Y.~H., \and Hirayama~Y.}
\REVIEW {Phys. Rev. Lett.}{91}{2003}{226804}.

\bibitem{Storcz2005a}
\Name{Storcz~M.~J., Hartmann~U., Kohler~S., \and Wilhelm~F.~K.}
\REVIEW {Phys. Rev. B}{72}{2005}{235321}.

\bibitem{Thorwart2005a}
\Name{Thorwart~M., Eckel~J., \and Mucciolo~E.~R.}
\REVIEW {Phys. Rev. B}{72}{2005}{235320}.

\bibitem{Makhlin2001a}
\Name{Makhlin~Y., Sch\"on~G., \and Shnirman~A.}
\REVIEW {Rev. Mod. Phys.}{73}{2001}{357}.

\bibitem{Mozyrsky2002a}
\Name{Mozyrsky~D., Kogan~S., Gorshkov~V.~N., \and Berman~G.~P.}
\REVIEW {Phys. Rev. B}{65}{2002}{245213}.

\bibitem{Mozyrsky1998a}
\Name{Mozyrsky~D. \and Privman~V.}
\REVIEW {J. Stat. Phys.}{91}{1998}{787}.

\bibitem{Yu2002a}
\Name{Yu~T. \and Eberly~J.~H.}
\REVIEW {Phys. Rev. B}{66}{2002}{193306}.

\bibitem{Yu2003a}
\Name{Yu~T. \and Eberly~J.~H.}
\REVIEW {Phys. Rev. B}{68}{2003}{165322}.

\bibitem{Reina2002a}
\Name{Reina~J.~H., Quiroga~L., \and Johnson~N.~F.}
\REVIEW {Phys. Rev. A}{65}{2002}{032326}.

\bibitem{Tolkunov2005a}
\Name{Tolkunov~D., Privman~V., \and Aravind~P.~K.}
\REVIEW {Phys. Rev. A}{71}{2005}{060308(R)}.

\bibitem{Braun2006a}
\Name{Braun~D.}
\REVIEW {Phys. Rev. Lett.}{96}{2006}{230502}.

\bibitem{Wootters1998a}
\Name{Wootters~W.~K.}
\REVIEW {Phys. Rev. Lett.}{80}{1998}{2245}.

\bibitem{Weiss1999a}
\Name{Weiss~U.}
\Book{Quantum Dissipative Systems},
\newblock 2nd edition
\Publ{World Scientific, Singapore}
\Year{1998}.

\bibitem{GradshteynRyzhik_ssq}
\Name{Gradshteyn~I.~S. \and Ryzhik~I.~M.}
\Book{Table of Integrals, Series, and Products},
\newblock 5th edition
\Publ{Academic Press, San Diego}
\Year{1994},
\newblock eq.~3.551(3).

\bibitem{Ferreira2006a}
\Name{Ferreira~A., Guerreiro~A., \and Vedral~V.}
\REVIEW {Phys. Rev. Lett.}{96}{2006}{060407}.

\end{thebibliography}
\end{document}